\documentstyle[prl,aps,multicol,epsf]{revtex}

\input{psfig.sty}
\renewcommand{\narrowtext}{\begin{multicols}{2}
\global\columnwidth20.5pc} 
\renewcommand{\widetext}{\end{multicols}
\global\columnwidth42.5pc} \multicolsep = 8pt plus 4pt minus 3pt

\begin{document}
\draft
\title{Polariton dynamics and Bose-Einstein condensation in semiconductor
microcavities}
\date{\today}
\author{D. Porras$^1$, C. Ciuti$^2$, J.J. Baumberg$^3$,
and C. Tejedor$^1$}
\address{$^1$ Departamento de F\'{\i}sica Te\'{o}rica de la Materia Condensada. 
Universidad Auton\'{o}ma de Madrid. 28049 Cantoblanco, Madrid, Spain.
\\$^2$ Physics Department, University of California San Diego.
 3500 Gilman Drive, La Jolla, CA 92093. USA.
\\$^3$ Department of Physics $\&$ Astronomy, University of Southampton,
Southhampton, SO17 IBJ, United Kingdom.
 }
\maketitle

\begin{abstract}
We present a theoretical model that allows us to describe the polariton dynamics in
a semiconductor microcavity at large densities, for the case of non-resonant
excitation. 
Exciton-polariton scattering from
a thermalized exciton reservoir is identified as the main mechanism for
relaxation into the lower polariton
states. 
A maximum in the polariton distribution that shifts towards lower energies
with increasing pump-power or temperature is shown, in agreement with recent
experiments. 
Above a critical pump-power, macroscopic occupancies ($5 \times 10^4$)
can be achieved in the lowest energy polariton state.
Our model predicts the
possibility of Bose-Einstein Condensation of polaritons, driven by
exciton-polariton interaction, at densities well below the saturation density
for CdTe microcavities.
\end{abstract}


\narrowtext

\section{Introduction}

Polaritons are quasiparticles created by the strong coupling between excitons
and photons, and behave as composite bosons at small enough densities. In the
last decade a huge experimental and theoretical effort has been dedicated to the
search for quantum degeneracy effects in microcavity polaritons.
They were soon proposed as candidates for the formation of a Bose-Einstein
condensate, due to their small density of states (DOS) \cite{Imamoglu}. 
However, a bottleneck effect due to the slow polariton-acoustical phonon
scattering was predicted \cite{Tassone97} and observed
in angle-resolved experiments \cite{Muller}.   
Stimulated scattering due to the bosonic nature of polaritons
has been demonstrated \cite{Huang} in pump-probe experiments. Parametric
amplification and parametric oscillation, under resonant excitation, 
has been observed \cite{Savvidis,Baumberg} and was well explained by
coherent polariton-polariton scattering \cite{Ciuti00,Ciuti01,Whittaker}.
In GaAs microcavities, experiments have shown the role of the 
polariton-polariton scattering
in the nonlinear emission after non-resonant excitation, either in continous wave
\cite{ring,Tartakovskii,Senellart}, or time-resolved experiments \cite{Martin}.
The growth of II-VI
microcavities leads to new exciting possibilities due to the larger stability of
the exciton. The study of these samples has allowed for the observation of
stimulated scattering
in the strong coupling regime under non-resonant excitation
conditions \cite{Dang}. The robustness of the polariton in these samples has
been recently shown by the demonstration of parametric amplification at high
temperatures \cite{Saba}. 

The theoretical study of semiconductor microcavities has followed two different
lines. A fermionic formalism, including carrier-carrier correlation, explained
the lasing by population inversion at densities above the saturation density,
where the coupling between the exciton and the photon disappears (weak coupling
regime) \cite{Kira,Khitrova}. On the other hand, a bosonic picture, including
polariton-polariton interaction, has been developed. It must be stressed that
recent experiments agree with the predictions made by the 
bosonic formalism
(bottleneck effect, stimulated scattering)  under
excitation conditions such that the density of polaritons remains below the
saturation density. The
polariton-polariton interaction, given by the usual Coulomb exchange interaction
between excitons, and the saturation term, has also explained quantitatively many 
recent observations \cite{Huang,Savvidis,Baumberg,Ciuti00,Ciuti01}.
   
In this paper we present a theoretical study of the evolution of the polariton
population at large densities with a bosonic description.
We use the semiclassical Boltzmann equation,
developed by Tassone {\it et al.}
\cite{Tassone}, as the starting point for the
description of the polariton dynamics.  
Exciton-polariton (X-P) scattering 
is identified as the most important
mechanism for relaxation towards the lower polariton branch. 
We deduce a simplified model in which the high energy excitons are considered as
a thermalized reservoir.
This simplification allows us to describe all the microscopic lower polariton 
states,
and the evolution of the polariton population towards
Bose-Einstein Condensation (BEC) \cite{Gardiner}. 

X-P scattering is shown to produce a polariton population with an occupancy that
has a maximum at a given energy below the exciton energy. The energy at which the
polariton distribution peaks is shown to decrease linearly as pump-power or sample
temperature is increased, 
in such a way that the dip in the polariton
dispersion behaves as a polariton trap in phase-space \cite{Baumberg01}.
This conclusion is valid for all semiconductor
microcavities. In particular, the same behavior has been observed in a recent
experiment in a GaAs microcavity \cite{ring}.
However, in GaAs, densities comparable to the saturation density are created, as
one increases the pump-power, before the threshold for BEC is reached. Thus,
for a clean observation of all the
phenomenology described in this paper, better material parameters than those of
GaAs microcavities are desirable.
We predict that in CdTe microcavities, where excitons
are more stable due to their smaller Bohr radius, the threshold for BEC can be
reached for densities well below the saturation density. Thus, we choose,
for concreteness, CdTe parameters for the detailed calculations shown below.
Our model shows that the relaxation towards the ground state is only possible for
densities comparable to the saturation density in GaAs microcavities.
In order to achieve BEC
with smaller polariton densities, an additional scattering channel that
those considered here should be
included. Electron-polariton scattering in doped microcavities has been
recently proposed \cite{Malpuech} as an efficient mechanism that allows BEC
for such small polariton densities. 

The paper is organized as follows. In section II we present the theoretical
framework. A set of differential equations is obtained for the description of
the microscopic lower polariton levels, and the exciton
density and exciton temperature. In section III, we apply this model to the study of the
dynamics of the polariton population for different pump-powers and
temperatures, after non-resonant, continuous excitation. The case of
Bose-Einstein Condensation into the lowest energy polariton mode is considered
in section IV. Finally, in section V, we list our main conclusions,
and discuss the comparison with experiments.

\section{Rate equation simplified model}

\subsection{Simplified model for the exciton-polariton scattering}

At densities such that $na_B^2 << 1$, where $a_B$ is the 2-D exciton Bohr
radius, excitons can be described as interacting bosons.
From typical parameters for CdTe QW's 
(exciton binding energy $E_B \approx$ 25 
$meV$, and electron and hole effective masses $m_e =$ 0.096, $m_h =$ 0.2),
we get
a 2-D exciton Bohr radius $a_B =$ 47 $\AA$, and  
$a_B^{-2} = 4.5 \times 10^{12} cm^{-2}$.
On the other hand, at large exciton densities, carrier exchange interactions and
phase-space filling effects destroy the coupling between excitons and photons. The
density at which the exciton oscillator strength is completely screened is the
saturation density, $n_{sat}$. Following 
Schmitt-Rink {\it et al.} \cite{Schmitt-Rink}, 
we can estimate, for a CdTe microcavity,
\begin{equation}
n_{sat} = \frac{0.117}{\pi a_{eff}^2} = 6.7 \cdot 10^{11} cm^{-2} ,
\label{nsat}
\end{equation}
where $a_{eff}^2 = a_B/2$. 
In InGaAs and GaAs QW's, $n_{sat}$ would result to be
$6.6 \times 10^{10} cm^{-2}$ and
$1.3 \times 10^{11} cm^{-2}$, respectively. 
Along this paper we consider the case of CdTe in 
a range of polariton densities
well below $n_{sat}$, so that the microcavity remains in the strong coupling
regime. 

In the strong coupling regime, the Hamiltonian includes the coupling between 
excitons and photons through the polariton splitting, $\Omega_P$ \cite{note}:
\begin{equation}
H_0 
= \sum_{\bf k} \left[
\epsilon^x_{\bf k} b^{\dagger}_{\bf k} b_{\bf k} 
+ \epsilon^c_{\bf k} a^{\dagger}_{\bf k} a_{\bf k} +
\frac{\Omega_P}{2} \left( a^{\dagger}_{\bf k} b_{\bf k} 
+ b^{\dagger}_{\bf k} a_{\bf k} \right) \right].
\label{H0}
\end{equation} 
$b^{\dagger}_{\bf k}$, $\epsilon^x_{\bf k}$ 
($a^{\dagger}_{\bf k}$, $\epsilon^c_{\bf k}$) 
are the
creation operator and dispersion relation for the bare exciton (photon) mode.
The
residual Coulomb interaction between excitons, and the saturation of the exciton
oscillator strength, can be described by the effective exciton-exciton, and
exciton-photon interaction Hamiltonian \cite{Hanamura}:
\begin{eqnarray}
H_I =  
\sum_{\bf{k_1},\bf{k_2} \atop \bf{q}} &&
\frac{M_{xx}}{2 S}  
 b^{\dagger}_{{\bf k}_1} b^{\dagger}_{{\bf k}_2} 
 b_{{\bf k}_1 + {\bf q}} b_{{\bf k}_2 - {\bf q}} +
\nonumber \\ 
& &  \frac{\sigma_{sat}}{S}  
 b^{\dagger}_{{\bf k}_1} b^{\dagger}_{{\bf k}_2} 
 b_{{\bf k}_1 + {\bf q}} a_{{\bf k}_2 - {\bf q}}
 + h.c.
\label{HI}
\end{eqnarray}
where $S$ represents the quantization area. 
We consider here the small momentum limit
for $H_I$, an approximation that is valid for the range of small wave-vectors
considered below. The matrix elements in $H_I$ are given by:
\begin{eqnarray}
M_{xx} \approx 6 \ E_B a_B^2
\nonumber \\
\sigma_{sat} \approx 1.8 \ \Omega_P a_B^2
\label{matrix}
\end{eqnarray}

\begin{figure}
\centerline{
\psfig{figure=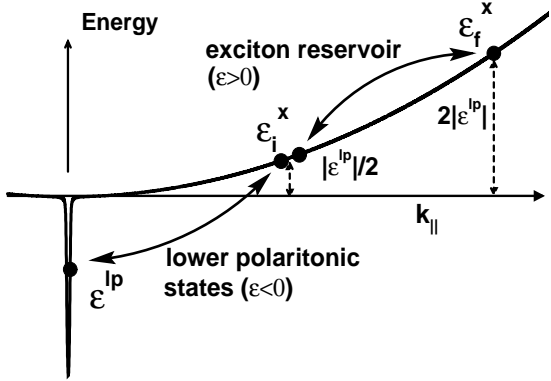,height=2.6in,width=3.0in}
}
\caption{Exciton-polariton scattering process from a thermalized exciton 
reservoir into the lower polariton branch.}
\label{inout}
\end{figure}

$H_0$ can be diagonalized in the polariton basis:
\begin{eqnarray}
p_{lp,\bf k} = X^{lp}_{\bf k} b_{\bf k} + C^{lp}_{\bf k} a_{\bf k} 
\nonumber \\ 
p_{up,\bf k} = X^{up}_{\bf k} b_{\bf k} + C^{up}_{\bf k} a_{\bf k}
\label{pbasis}
\end{eqnarray}
where $X^{lp/up}_{\bf k}$, $C^{lp/up}_{\bf k}$ are the Hopfield coefficients, 
that represent the
excitonic and photonic weights in the polariton wave-function
for the lower/upper branch, respectively
\cite{Hopfield}. By means of the transformation (\ref{pbasis})
we can express $H_I$
in terms of polariton operators. We will neglect the upper polariton branch,
an approximation that will be justified later. We express the operators 
$b_{\bf k}$, $a_{\bf k}$ in (\ref{HI}) in terms of 
$p_{lp, \bf k}$, $p_{up, \bf k}$, and
retain the terms corresponding to interactions within the lower polariton branch
only, 
\begin{eqnarray}
H &=& \sum_{\bf k} \epsilon_{\bf k}^{lp} p_{lp,\bf k}^{\dagger} p_{lp,\bf k}^{}
\nonumber \\
+ && \sum_{\bf{k_1},\bf{k_2} \atop \bf{k_3},\bf{k_4}} 
\frac{1}{2 S} V_{\bf{k_1},\bf{k_2},\bf{k_3},\bf{k_4}}^{lp-lp}
p_{lp,\bf{k_1}}^{\dagger} p_{lp,\bf{k_2}}^{\dagger} 
p_{lp,\bf{k_3}}^{} p_{lp,\bf{k_4}}^{},
\label{hamiltonian}
\end{eqnarray}
where 
\begin{eqnarray}
\frac{1}{2} V_{{\bf k}_1,{\bf k}_2,{\bf k}_3,{\bf k}_4}^{lp-lp} &&= \nonumber \\ 
\left( \frac{1}{2} M_{xx} \right. && 
   X_{\bf k_1}^{lp} X_{\bf k_2}^{lp} X_{\bf k_3}^{lp} X_{\bf k_4}^{lp} +
 \nonumber \\ 
   \sigma_{sat} && 
   C_{\bf k_1}^{lp} X_{\bf k_2}^{lp} X_{\bf k_3}^{lp} X_{\bf k_4}^{lp} + 
 \nonumber \\
   \sigma_{sat} && \left. 
   X_{\bf k_1}^{lp} X_{\bf k_1}^{lp} C_{\bf k_3}^{lp} X_{\bf k_4}^{lp} \right)
\delta_{{\bf k}_1 + {\bf k}_2, {\bf k}_3 + {\bf k}_4} .
\label{interaction}
\end{eqnarray}
$\epsilon_{\bf k}^{lp}$ is the dispersion relation for the lower polariton branch.
It is plotted, assuming typical parameters for a semiconductor microcavity, in
Fig. \ref{inout}. For
energies higher than the bare exciton energy, it matches the bare exciton
dispersion.

The theoretical model 
developed by Tassone et {al.} \cite{Tassone} describes the evolution of
the polariton population by means of a 
semiclassical Boltzmann equation that includes
polariton-acoustical phonon
and polariton-polariton scattering. 
A grid uniform in energy is considered for the description of the polaritonic
levels.
Here, we are interested in
the evolution of the system towards a Bose-Einstein condensate. It is well
known that in 2-D systems, BEC is possible at finite temperatures, only for
systems with a finite quantization length, $L_c$. As a result, 
the critical density and
temperature depend logarithmically on this length scale. 
Due to the peculiarity introduced
by the 2-D character of semiconductor microcavities,
the grid uniform
in energy considered in \cite{Tassone} is not well suited for the study of 
the transition towards BEC. 
Instead, a uniform grid in k-space, related with the inverse of the quantization
length, $L_c^{-1}$, has to be considered. Unfortunately,
the number of levels that one has to consider for the
description of the polariton dynamics is dramatically increased,
and a numerical calculation that includes all the possible polariton-polariton 
scattering
processes is not possible. Thus, some simplification has to be done, which
still allows to describe satisfactorily the polariton dynamics, 
and to predict the
possibility of Bose-Einstein Condensation.

In order to develop a simplified model for the case of non-resonant excitation, 
we make the following assumptions:

(1) The exciton population above the bare exciton energy can be
considered as a reservoir at a given temperature, that follows
a Maxwell-Boltzmann distribution.

(2) The main scattering mechanism at large densities 
for relaxation into the lower polariton modes is the one shown in Fig.
\ref{inout}, in which two excitons scatter in the reservoir, and the final states
are a lower polariton, and another reservoir exciton. We will call this process
exciton-polariton scattering.

A scattering process in which two excitons are initially in the exciton reservoir,
and a lower and upper polariton are the final states is also possible, but has a
much smaller probability. This is due to the fact that in such a collision, an
upper polariton, instead of an exciton, plays the role of a final state. Thus,
the scattering rate is reduced by 
the ratio between the upper polariton and bare exciton DOS,
$\rho_{up}/\rho_x \approx 4 \times 10^{-5}$. In fact, for negative 
or zero detunings, the upper polariton population
remains negligible, as shown experimentally \cite{Dang} and theoretically
\cite{Tassone}. The scattering of excitons with acoustical phonons going 
into lower polariton states is also much slower than the X-P scattering 
already at exciton densities of $10^9 cm^{-2}$, and the lattice temperatures
considered here. 

The phase-space is, thus, divided into two pieces.

\begin{itemize}

\item 
{\it The exciton reservoir}, which
describes the excitons directly
created by the non-resonant pump. We take as 
zero energy the bare exciton energy, so that the exciton reservoir includes all
the levels with $\epsilon > 0$. 
These levels are weakly coupled to the photonic modes, so that they can be
approximated as bare excitons.
The occupation numbers are given by,
\begin{equation}
N_i^x = N^x e^{-\epsilon_i^x/k_B T_x} .
\label{MB}
\end{equation}
$N^x$ is given by
\begin{equation}
N^x = \frac{2 \pi n_x}{\rho_x k_B T_x} ,
\label{Nx}
\end{equation}
where $n_x$ is the exciton density, $\rho_x$ is the bare exciton density of
states,
\begin{equation}
\rho_x = (m_e+m_h)/\hbar^2 ,
\label{dos}
\end{equation}
and $T_x$ is the temperature of the exciton reservoir. The assumption of a
rigid thermal exciton distribution is well justified, as discussed below,
for levels above $\epsilon = 0$, due to the fast exciton-exciton scattering
that thermalize the
exciton population at large densities \cite{Tassone,Ivanov}.
On the other hand, a Maxwell-Boltzmann distribution, rather than a Bose-Einstein
distribution can be used to describe the exciton reservoir, provided that $N^x <<
1$. $N^x \approx 0.2$ for the largest densities considered here, so that
quantum degeneracy effects are negligible.
The main mechanism
for radiative losses in the exciton reservoir 
is due to the coupling to the leaky modes, that we
include by considering an exciton lifetime, $\tau_x = 100ps$
\cite{Tassone96}.

\item 
{\it The lower polariton branch (LPB)}, where the energy levels are below the bare
exciton energy ($\epsilon_k^{lp} < 0$).  
Exciton and photon levels in the LPB
are strongly coupled, and radiative losses are much more important, due to the
photonic weight of these states. Thus, we expect a nonequilibrium polariton
distribution that depends strongly on temperature and density. 
On the other hand, the abrupt dispersion
relation implies a very small effective mass
for the LPB states, 
$m_{lp} \approx 4 \times 10^{-5} m_x$,
leading to the possibility of Bose-Einstein Condensation.  

\end{itemize}

With this separation of the phase-space we are neglecting the details of the
polariton population near $\epsilon \approx 0$, that is, at the knee in the polariton
dispersion relation,
where the LPB and the quadratic dispersion of the bare excitons merge
together (see Fig. \ref{inout}). The contribution from these states is 
negligible already at
the lowest temperatures considered here ($4K$, $k_B T \approx 0.35 meV$),
because the main part of the exciton distribution is in the weak coupling
region.
We describe the LPB states by a grid in k-space.
The exact quantization area is not well determined in
experiments, due to inhomogeneities in the spot, and surface defects. The relevant
quantity is the quantization length, that gives the order of magnitude for the
distance between the energy levels. 
We consider first $L_c = 50 \mu m$ (a typical excitation spot diameter),
and plane waves  for the microscopic polariton states, so that the quantization
area is simply $S = L_c^2$. The allowed 
wave-vectors are given by
\begin{equation}
{\bf k}^{lp} = (n_{\bf x} {\bf x} + n_{\bf y} {\bf y}) \frac{2 \pi}{L_c} ,
\label{levels}
\end{equation}
where $n_{\bf x}, n_{\bf y}$ are integers.
A similar method has been employed to describe the microscopic polariton levels
in a microcavity etched into a microscopic post structure ($L_c = 2 \mu m$)
\cite{Tassone00}. In
that case, the LPB is reduced to a single microscopic level. On the contrary, in
our case, $L_c$ is larger, and the number of lower polariton levels results to
be of the order of $10^3$.  
The main results of our work do not depend
strongly on the exact value of $L_c$, 
but only on its order of magnitude, as we show in the last section. The weak
dependence of our results on $L_c$ can be easily understood from the peculiarities
of BEC in 2-D (see Appendix B). 

We write now a set of rate equations for $N_k^{lp}$ 
(occupation numbers in the lower polariton
branch) and $n_x$ (the density of the thermalized exciton reservoir).
We consider the injection of polaritons from isotropically distributed excitons,
so that the occupation numbers $N_k^{lp}$ will depend on the absolute value
of the in-plane wave-vector only.
The rates for scattering
from the thermalized reservoir into and out of the LPB, due to the interaction
Hamiltonian (\ref{HI}) are calculated in Appendix A. 
They lead to a simplified version of the semiclassical Boltzmann equation:
\begin{eqnarray}
\frac{dN_k^{lp}}{dt} &=&
\nonumber \\ 
& & W^{in}_k n_x^2 (1+N_k^{lp}) 
        - W^{out}_k n_x N_k^{lp} - \Gamma_k^{lp} N_k^{lp}
\nonumber \\
\frac{dn_x}{dt} &=& 
\nonumber \\
& &- \frac{1}{S} \sum_{k} 
dg_k^{lp} ( W^{in}_k n_x^2 (1+N_k^{lp}) - W^{out}_k n_x N_k^{lp} )
\nonumber \\
&-& \Gamma_x n_x + p_x  .
\label{model}
\end{eqnarray} 

$\Gamma_k^{lp}$, $\Gamma_x$ are the radiative losses in the LPB and exciton
reservoir, respectively:
\begin{eqnarray}
\Gamma_k^{lp} &=& |C^{lp}_k|^2 \frac{1}{\tau_c}  \nonumber \\
\Gamma_x &=& \frac{1}{\tau_x} ,
\label{radiativelosses}
\end{eqnarray}
where $\tau_c$ is the lifetime for the bare photon within the cavity. We
consider $\tau_c = 1ps$, a typical value for high quality microcavities. $p_x$
represents the non-resonant pump that injects excitons directly into the exciton
reservoir. If relaxation into the LPB is slow enough, the lifetime for an
exciton in the exciton reservoir is given by $\tau_x$, so that a given pump
$p_x = n_x/\tau_x = n_x/100 ps$
creates an exciton density $n_x$. If scattering into
the LPB is fast enough, however, the density created by the pump is smaller, as we
will see below.
$dg_k^{lp}$ is the degeneracy for the polariton level corresponding to the
wave-vector $k$, and is calculated according to the distribution obtained from
Eq. (\ref{levels}).

$W_{k}^{in} n_x^2$, $W_{k}^{out} n_x$ are the rates for X-P scattering
into and out of the LPB. They are given by the expressions (see Appendix A):
\begin{eqnarray}
W^{in}_k &=& \frac{2 \pi}{\hbar k_B T_x} M_k^2 e^{\epsilon^{lp}_k/k_B T_x}
\nonumber \\
W^{out}_k &=& \frac{1}{\hbar} M_k^2 \rho_x e^{2 \epsilon^{lp}_k/k_B T_x} ,
\label{rates}
\end{eqnarray}
where 
\begin{equation}
M_k = M_{xx} X^{lp}_k + \sigma_{sat} C^{lp}_k .
\label{matrixrates}
\end{equation}

The dependence of the rates on temperature can be qualitatively understood by 
the phase-space restrictions for relaxation into and out of the LPB
(Fig. \ref{inout}). Let us consider two excitons
in the exciton reservoir with a wave-vector 
$k_i^x$ and energy $\epsilon_i^x$, that
scatter to a LPB state with energy $\epsilon_k^{lp}$ and to a high energy state
with $k_f^x$, $\epsilon_f^x$.
Taking into account that the wave-vector of the low-energy
polariton is negligible, when compared with the wave-vectors in the reservoir,  
simultaneous energy and momentum conservation implies that: 
\begin{eqnarray}
2 \ |k_i^x|=|k_f^x| &\rightarrow& 4 \ \epsilon_i^x = \epsilon_f^x 
\nonumber \\
2 \ \epsilon_i^x = - |\epsilon_k^{lp}| + \epsilon_f^x &\rightarrow& \epsilon_i^x 
= |\epsilon_k^{lp}|/2, \ \ \epsilon_f^x = 2 \ |\epsilon_k^{lp}| .
\label{condition}
\end{eqnarray}   
Eq. (\ref{condition}) means that two excitons need an energy 
$|\epsilon_k^{lp}|/2$ to be the
initial states of a process in which one of them falls into the LPB.
On the other hand, an exciton needs an energy 
$2 \ |\epsilon_k^{lp}|$ in order to
scatter with a lower polariton and take it out of the LPB. 

The polariton distribution in the steady-state regime will be
given by the set of equations (\ref{model}) with
$dN_k^{lp}/dt = 0$, $dn_x/dt = 0$.
In the steady-state, the occupation numbers in the LPB are given
by:
\begin{equation}
N^{lp}_k = \frac{W_k^{in} n_x^2}{W_k^{out} n_x 
+ \Gamma_k^{lp} - W_k^{in} n_x^2} .
\label{steady}
\end{equation}  
If we neglect the dependences on $k$ 
of the matrix element $M_k$ and the radiative
losses $\Gamma_k^{lp}$, then the maximum in the polariton distribution has a 
peak at the energy:
\begin{equation}
|\epsilon_{max}^{lp}| = \frac{1}{2} k_B T_x log \left( \frac{1}{\hbar \Gamma^{lp}}
 M^2 \rho_x n_x  
 \right) . 
\label{energyloss}
\end{equation}
Eq. (\ref{energyloss}) implies an approximately linear relation between the
exciton reservoir temperature and the energy position at which the system
prefers to relax into the LPB (that is, the maximum energy
loss in the relaxation process).

\subsection{Model for the evolution of the exciton reservoir temperature}

The strong dependence of the X-P scattering into the LPB on 
$T_x$ implies that, in order to complete our model, we have to 
describe properly the evolution of the exciton reservoir temperature.
During the 
relaxation process into the LPB, shown in Fig. \ref{inout}, a
high energy exciton is injected into the exciton reservoir. This way, during the
relaxation process, the exciton reservoir is heated. 
On the other hand, the scattering with acoustical phonons, tries to
keep the exciton reservoir at the lattice temperature.
In order to describe the evolution of the exciton reservoir temperature, a new
variable has to be added to this set of equations. We introduce $e_x$, the
density of energy for the exciton reservoir, that is given, for a
Maxwell-Boltzmann distribution, by the expression:
\begin{equation}
e_x = \frac{1}{S}\sum_k N_k^x \ \epsilon_k^x = n_x (k_B T_x) ,
\end{equation}
The evolution of the energy allows us to obtain the exciton temperature, at each
step in the evolution of the polariton population. 

However, we still need to make an assumption about the exciton distribution that
the pump directly injects into the reservoir. Non-resonant experiments usually
create a population of high energy electron-hole pairs, that relax towards low
energy states, after the process of exciton formation 
\cite{Tassone97,Piermarocchi}. It
must be stressed that for the case of CdTe QW's this process is particularly fast,
due to the possibility of exciton formation through the emission of LO phonons,
whose energy is 21 meV, below the exciton binding energy (25 meV). The time scale
for thermalization due to exciton-exciton scattering is very fast ($\approx
1ps$), as compared with the time scale for relaxation into the LPB ($\approx 20
ps$, for the larger exciton densities considered here). Thus, it is 
very well justified
to describe the non-resonant pump, as a source that injects thermalized excitons at
a given pump-temperature into the exciton reservoir. We still need to specify which
is the initial pump-temperature. Due to the large uncertainties in non-resonant
experiments it is very difficult to make predictive calculations about this initial
condition. We choose instead to make the simplest assumption: we consider that the
pump injects excitons at the lattice temperature, $T_L$. For large exciton densities
(above $5 \times 10^{10} cm^{-2}$) the processes of reservoir heating by X-P
scattering and cooling through exciton-phonon scattering, govern the final exciton
temperature, and our results do not depend on our assumption for the
pump-temperature. For small temperatures (below 10 K), and densities, our results
depend on this assumption. In a real experiment, the exciton temperature can
be larger than the one considered here, if the cooling of the initial
non-resonantly created excitons is not fast enough.

It must be pointed out that a
very efficient experimental setup for non-resonant excitation of a cold
exciton distribution in a semiconductor
microcavity has been demonstrated by Savvidis {et al.} \cite{ring}. In this
experiment, non-resonant excitation has been achieved with small excess energies
(below the stop band), due to the finite transmittance of the Bragg reflectors.
The experimental results presented in \cite{ring} are in agreement with a model
in which excitons relax from a thermalized distribution into the lowest energy
states, and justify all the assumptions made above. 
   
We have to quantify the three contributions for the evolution of the exciton
temperature listed above:

\vspace{0.1in}  
{\it(i)  Heating by the scattering to high energy exciton levels.} 
\vspace{0.1in} 

The total energy is conserved in the X-P scattering towards the 
LPB. Thus, the energy lost by the polaritons falling into the lower energy
states must be gained by the exciton reservoir. The following expression gives
the rate of change of $e_x$ by each X-P
scattering process:
\begin{eqnarray}
\frac{de_x}{dt} \mid_{X-P} &=& - \frac{1}{S} \sum_k \epsilon_k^{lp} dg_k^{lp}
\nonumber \\
&\times& (
W^{in}_k n_x^2 (1+N_k^{lp}) - W^{out}_k n_x N_k^{lp}) .
\label{XPheating}
\end{eqnarray}

The heating of the exciton gas in the relaxation towards the LPB was also shown
theoretically in \cite{Tassone00}, for the case of a microcavity post.

\vspace{0.1in} 
{\it (ii) Cooling produced by acoustical phonons.}
\vspace{0.1in} 

The exciton-phonon scattering tries to keep excitons at the lattice
temperature, $T_L$, against the heating considered above \cite{Ivanov}.
We divide the exciton Maxwell-Boltzmann distribution into levels equally spaced
in energy with a given $\Delta E << k_B T_L, k_B T_x$. The probability for
absorption or emission of acoustical phonons gives a scattering rate 
$W^{ph}_{j \rightarrow i} (T_L)$ for the transition between states with energies
$\epsilon_i^x, \epsilon_j^x$. These rates depend on the phonon distribution
through $T_L$, and can be evaluated from the
exciton-acoustical phonon interaction Hamiltonian.
The
evolution of the occupation numbers, $N_i^x$, in the exciton reservoir leads 
to an expression for the rate of change of the total energy:
\begin{eqnarray}
& & \frac{de_x}{dt} \mid_{ph} = \frac{\rho_x}{2 \pi} \sum_i
 \Delta E \ \epsilon_i^x \frac{d N_i^x}{dt} \mid_{ph}=
\nonumber \\
&=& \frac{\rho_x \Delta E}{2 \pi} \sum_i \epsilon_i^x \sum_j \left( 
W^{ph}_{j \rightarrow i} (T_L) N_j^x 
- W^{ph}_{i \rightarrow j}(T_L) N_i^x \right)
\nonumber \\
&=& \frac{\Delta E \ n_x }{k_B T_x} \sum_i \epsilon_i^x \times
\nonumber \\ 
& & \sum_j  \left(  
W^{ph}_{j \rightarrow i} (T_L) e^{-\epsilon_j^x/k_B T_x} 
- W^{ph}_{i \rightarrow j}(T_L) e^{-\epsilon_i^x/k_B T_x} \right) .
\nonumber \\
\label{Phononcooling}
\end{eqnarray} 

For the calculation of $W^{ph}$ we have used the exciton-acoustical phonon
coupling through the deformation potential interaction. These rates have been
presented in several previous works \cite{Tassone,Ivanov,Piermarocchi}, and we
will not re-write them here. 
We consider deformation potentials $a_e= 2.1 eV$,
$a_h=5.1 eV$ for CdTe \cite{deformation}, and an $80 \AA$  QW. 
In Eq. (\ref{Phononcooling}), stimulated terms are neglected, 
due to the small occupation
numbers in the exciton reservoir. The rate of change for $e_x$ results to be
linear in $n_x$, and dependent on both $T_L$ and $T_x$.

\vspace{0.1in} 
{\it (iii) Cooling to the pump temperature.}
\vspace{0.1in}

The injection of excitons into the reservoir by a pump that is considered to be at
the lattice temperature tries to keep the exciton reservoir at $T_L$. The
terms that correspond to pumping and radiative losses in Eqs.
(\ref{model}),
\begin{equation}
\frac{dn_x}{dt} \mid_{pu} = p_x -\Gamma_x n_x ,
\label{modelpump}
\end{equation} 
imply that the rate for the evolution of the occupation numbers has the
contribution:
\begin{equation}
\frac{d N_k^x}{d t} \mid_{pu} = 
\frac{2 \pi p_x}{\rho_x k_B T_L} e^{- \epsilon_k^x /k_B T_L} -
\Gamma_x \frac{2 \pi n_x}{\rho_x k_B T_x} e^{- \epsilon_k^x /k_B T_x} ,
\label{modelpump2}
\end{equation}
where the condition that the pump injects excitons at $T_L$ into a
Maxwell-Boltzmann distribution at $T_x$ is included. The rates in Eq.
(\ref{modelpump2}) leads to a new contribution to the evolution of $e_x$:
\begin{equation}
\frac{de_x}{dt} \mid_{pu} = 
\frac{1}{S}
\sum_k \epsilon_k^x \frac{d N_k^x}{dt} \mid_{pu} =
p_x k_B T_L - \frac{n_x}{\tau_x} k_B T_x .
\label{Pumpcooling}
\end{equation}
Note that this contribution cools the exciton reservoir if $T_x > T_L$. 

The set of first-order differential equations (\ref{model}),
together with
(\ref{XPheating}, \ref{Phononcooling}, \ref{Pumpcooling}),
leads to a closed description for the
dynamics of the exciton reservoir and lower polaritons.
We have integrated numerically these equations until the stationary solution 
($dN^{lp}/dt = dn_x/dt = de_x/dt = 0$) is reached. The obtained solution
describes the steady-state for the case of a continuous non-resonant pump.

\subsection{Comparison between our simplified model and the complete
semiclassical Boltzmann equation.}

\begin{figure}
\centerline{
\psfig{figure=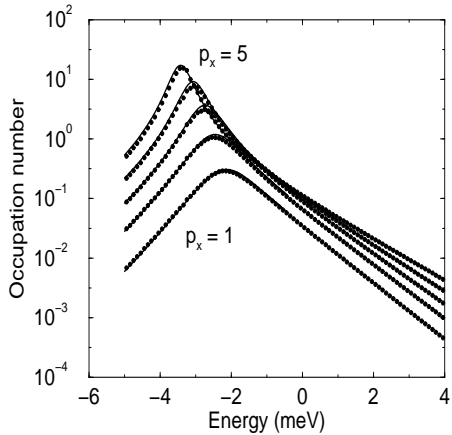,height=2.5in,width=2.5in}
}
\caption{Polariton distribution calculated by means of the complete 
semiclassical Boltzmann equation, as in Ref. [18] (points),
and  polariton distribution calculated by means of our
simplified model (continuous line). We have
considered a CdTe microcavity with zero detuning and $T_L = 10 K$. 
The curves, from bottom to top, refer to pump-powers from $1$ to $5$, in units of
$10^{10} cm^{-2}/100 ps$.}
\label{comparison}
\end{figure}

We have checked whether the 
polariton distribution obtained with
this simplified model agree
with the one obtained in a detailed calculation performed with a
Boltzmann equation in which the simplifying assumption of a Maxwell-Boltzmann
distribution for the levels in the exciton reservoir is not made. 
In Fig. \ref{comparison} we present the comparison between a complete
calculation that follows reference \cite{Tassone}, 
with a grid uniform in energy ($\Delta E = 0.1 meV$),
in which all the polariton-polariton and polariton-phonon scattering processes
are included, and
our simplified model. We expect that the results from both 
calculations coincide
under conditions for which Bose-Einstein Condensation (that is, the macroscopic
occupancy of a single microscopic mode) is not possible. On the contrary, in the
range of densities near or above BEC, comparison between both models would not
be allowed because the system would not be free to relax into a single 
microscopic mode.
Thus, a continuum density of states would not be able to describe properly 
the LPB, and a grid uniform in energy could not be applied.
In Fig. \ref{comparison} a few pump-powers have been chosen, for which the 
comparison is possible.
It can be clearly seen that both calculations yield identical
results, so that, we can conclude that
our simplified model captures the essential physics in the
relaxation from the exciton reservoir towards the lower energy polariton states.

\section{Evolution of the polariton distribution with pump-power and temperature}

\subsection{Evolution with temperature}

In the previous section, we obtained an estimation for the evolution of the
polariton population as the exciton temperature $T_x$ is increased.
A rough linear relation between the position of the maximum in the distribution
and $T_x$ was predicted.
Now we present a
complete calculation with our model that confirms this estimate. 
This effect can
be observed for a different range of polariton splittings and detunings. 
However, the
main parameter that governs the relaxation from the exciton reservoir
is the energy difference between the bare exciton and the bottom of the LPB.
This energy distance can be interpreted as the depth of the polariton trap
formed by the dip in the polariton dispersion. 
Within this paper we will consider, for concreteness, the case of zero detuning (that
is, $E_x$ (bare exciton energy) $=$ $E_c$ (bare photon energy)), so that the
lowest polariton mode has an energy $\Omega_P/2$ below $E_X$. 

\begin{figure}
\centerline{
\psfig{figure=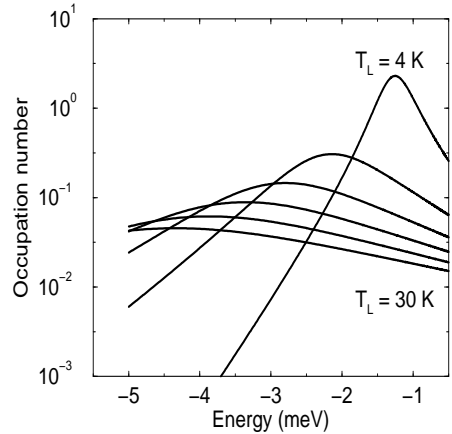,height=2.5in,width=2.5in}
}
\caption{Evolution of the polariton distribution for $T_L = 
4,10,15 \dots 30 K$,
$\Omega_P=$ 10 meV, zero detuning, and 
$p_x = 10^{10} cm^{-2}/100 ps$.}
\label{peakTemp10}
\end{figure}

\begin{figure}
\centerline{
\psfig{figure=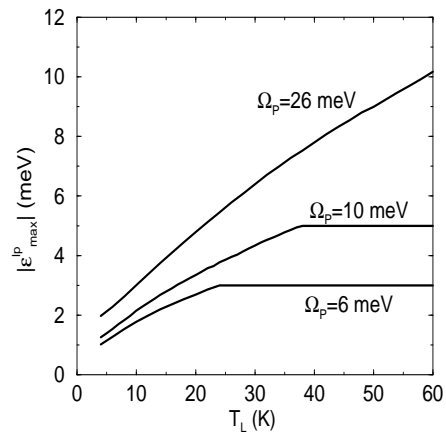,height=2.5in,width=2.5in}
}
\caption{Dependence of the energy of the 
peak in the polariton distribution on $T_L$ for
$p_x = 10^{10} cm^{-2}/100 ps$, and different polariton splittings.}
\label{maxTemp}
\end{figure}

In Fig. \ref{peakTemp10} we show the evolution of the polariton distribution for
$\Omega_P = 10 meV$. We choose $p_x=10^{10} cm^{-2} / 100 ps$.
Under such excitation conditions, X-P scattering is slower 
than radiative recombination
in the exciton reservoir, and $n_x \approx
\tau_x p_x$. The heating of the exciton gas is also negligible, so that $T_x
\approx T_L$. The peak in the polariton distribution moves towards lower
energies as temperature is increased. Due to the fact that the rate
given in Eq. (\ref{rates}) for scattering into the LPB decreases with $T_x$,
smaller occupation numbers are predicted for higher temperatures. 
At high $T_x$, the bottleneck is suppressed and the polariton distribution becomes
almost flat.
The difference
between high and low $T_x$ was already observed in \cite{Tartakovskii}, where a
flat distribution was evidenced for 
pump-powers where the X-P scattering is the dominant
relaxation mechanism.
The evolution of the peak in the polariton distribution is shown in 
Fig. \ref{maxTemp} for a range of polariton splittings.
The linear dependence predicted by the estimation in
Eq.(\ref{energyloss}) is clearly observed. For high enough temperatures and small
energy distance between the lowest polariton energy and the exciton reservoir,
the maximum in the polariton distribution can reach the bottom of the LPB.
However, the
occupation numbers are still smaller than one,
and the
situation is not able to produce BEC.
Due to the isotropy of the lower polariton distribution, the emission at a given
energy 
$\epsilon^{lp}_{max}$,
corresponds to emission at a given angle with the growth axis.
The movement of $\epsilon^{lp}_{max}$ 
towards lower energies implies that the sample emits light
preferentially in a ring, at a given angle that decreases with temperature. 
The recent experiment reported by Savvidis {et al.} \cite{ring} has observed the
ring emission with the linear dependence on lattice temperature described above,
and shows that the X-P scattering from a thermalized exciton reservoir
considered here, is the main scattering mechanism for relaxation into the LPB.

\subsection{Evolution with Pump-Power}

As shown in Eq. (\ref{energyloss}),
$\epsilon_{max}^{lp}$ depends logarithmically
on the exciton density, and linearly on $T_x$.
At large densities, however, the heating of the exciton reservoir must be
properly taken into account. When occupation numbers $N^{lp}_k>1$ are achieved,
the X-P scattering mechanism shown in Fig. \ref{inout} is stimulated, and 
the injection of high energy excitons becomes
more and more important. In this situation, the heating of the exciton reservoir
governs the evolution of $\epsilon_{max}^{lp}$ as $p_x$ is
increased, because for higher $T_x$, the polariton distribution has a maximum at
lower energies. This way, the heating of the exciton reservoir allows the
excitons to fall into the polariton trap formed by the dip in the polariton
dispersion.  

\begin{figure}
\centerline{
\psfig{figure=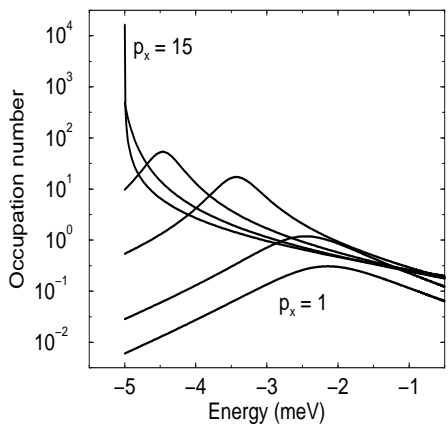,height=2.5in,width=2.5in}
}
\caption{Evolution of the polariton distribution for $T_L=10K$,
$\Omega_P=10meV$, 
and $p_x=$ $1,2,5,8,$ $15\times 10^{10} cm^{-2}/100 ps$ 
(from bottom to top).}
\label{peakPump10}
\end{figure}

\begin{figure}
\centerline{
\psfig{figure=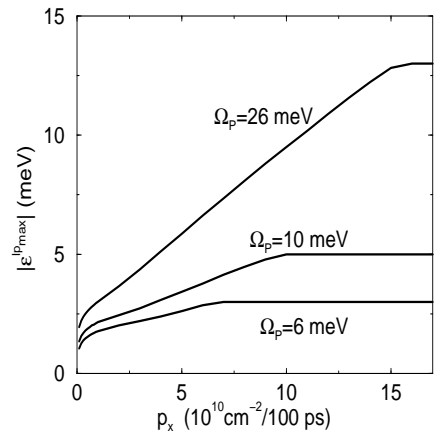,height=2.5in,width=2.5in}
}
\caption{Dependence of the energy of the peak in the polariton distribution 
on pump-power, for $T_L = 10 K$, and different polariton splittings.}
\label{maxPump}
\end{figure}

\begin{figure}
\centerline{
\psfig{figure=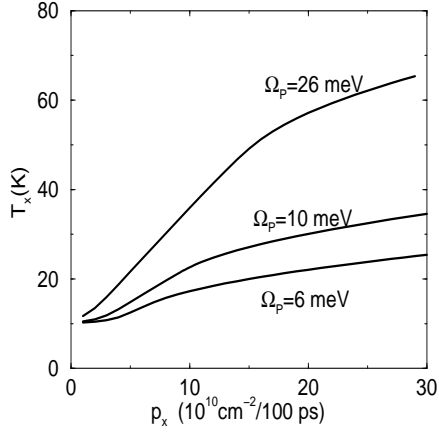,height=2.5in,width=2.5in}
}
\caption{Dependence of the temperature of the exciton reservoir 
on pump-power for $T_L=10 K$, and different polariton splittings.}
\label{tempEvol}
\end{figure}

\begin{figure}
\centerline{
\psfig{figure=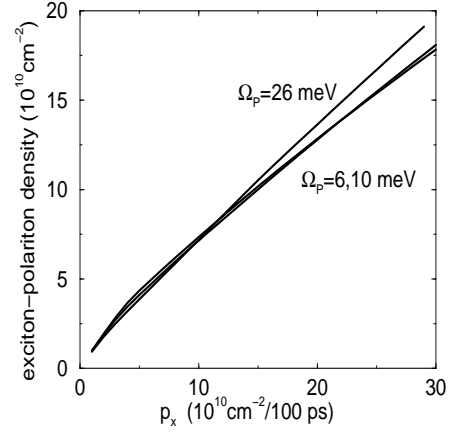,height=2.5in,width=2.5in}
}
\caption{Dependence of the total exciton reservoir and lower polariton density,
$n_{xp}$, on pump-power, for $T_L = 10 K$. The results for $\Omega_P =6$, and
$10 meV$ are indistinguishable.}
\label{denEvol}
\end{figure}

Fig. \ref{peakPump10} shows the evolution of the polariton
distribution as one increases the pump-power. The distribution can reach the
lowest energy state and large occupation numbers can now be achieved. For large
enough powers, the lowest energy state can have the main part of the polariton
population, a situation that will be described in detail in section IV. 
For intermediate pump-powers, occupation numbers larger than one are obtained for
modes with $k \neq 0$. Thus, our model predicts the stimulation towards these
modes before BEC is achieved.
In Fig. \ref{maxPump} we show the evolution of $\epsilon^{lp}_{max}$, as one
increases $p_x$, for different $\Omega_P$'s.
Three stages can be clearly distinguished.
From $p_x \approx 10^9 cm^{-2}/100ps$ to $p_x \approx 10^{10} cm^{-2}/100ps$,
the exciton density
created by the pump is small enough for the heating of the exciton reservoir to
be neglected. Under such conditions, the evolution of $\epsilon^{lp}_{max}$ is
governed by the logarithmic dependence on $n_x \approx p_x \tau_x$ (see Eq.
\ref{energyloss})). 
Above $p_x \approx 10^{10} cm^{-2}/100 ps$, the heating of the exciton gas is
important, and the evolution of $\epsilon^{lp}_{max}$ is governed by the
evolution of $T_x$, through the linear dependence shown in Eq.
(\ref{energyloss}). In this regime, $\epsilon^{lp}_{max}$ results to depend
linearly on $p_x$, until it saturates to the bottom of the LPB. 
The heating of the exciton reservoir is shown in Fig. \ref{tempEvol}. For large
pump-powers excitons are no longer at the temperature $T_L$,
at which they are initially injected. The raise of $T_x$ becomes more important
for larger $\Omega_P$ due to the highest energies of the excitons
injected into the reservoir by the X-P scattering (see Fig. \ref{inout}).
At large $p_x$, X-P scattering is faster than the radiative recombination given
by $\tau_x$, an it becomes the main mechanism for losses in the 
exciton reservoir.
Thus, the 
exciton-polariton density, $n_{xp}$ (that is, the density of excitons in the
reservoir plus the density of lower polaritons), 
created by the pump is smaller than $p_x
\tau_x$. Fig. \ref{denEvol} shows the evolution of the steady-state polariton
density as $p_x$ is increased. For all the three polariton splittings considered the
evolution of the exciton-polariton density is very similar.

Fig. \ref{peakPump10} and Fig. \ref{denEvol} allow to estimate the $n_{xp}$ 
at which the relaxation
into the ground state is possible. In Fig. \ref{peakPump10} pump-powers above $5
\times 10^{10} cm^{-2}/100 ps$ correspond to polariton densities larger than $4
\times 10^{10} cm^{-2}$. Thus, in GaAs microcavities, the weak coupling regime will
be reached before $\epsilon^{lp}_{max}$ reaches the bottom of the LPB dispersion,
due to the small $n_{sat}$, and BEC is not possible. 
For the clean observation of the complete evolution of
the
polariton distribution towards the ground state, CdTe microcavities are promising
candidates because $n_{xp}$ remains well below $n_{sat}$, before the
system relaxes towards the ground state.
The shift of $\epsilon^{lp}_{max}$ towards lower energies
results also in the emission in a ring at an angle that decreases as pump-power is
increased. In the experiments performed by Savvidis {et al.} \cite{ring}, this
effect was observed, and a linear relation between 
$\epsilon^{lp}_{max}$ and
pump-power was reported. 
In this experiment, the extrapolation of the linear dependence of
$\epsilon^{lp}_{max}$ on $p_x$ towards small powers shows an energy offset, in
agreement with the theoretical result (see Fig. \ref{maxPump}). 
Angle-resolved measurements allow, thus, to monitor the
evolution of the polariton population towards the ground state.

\begin{figure}
\centerline{
\psfig{figure=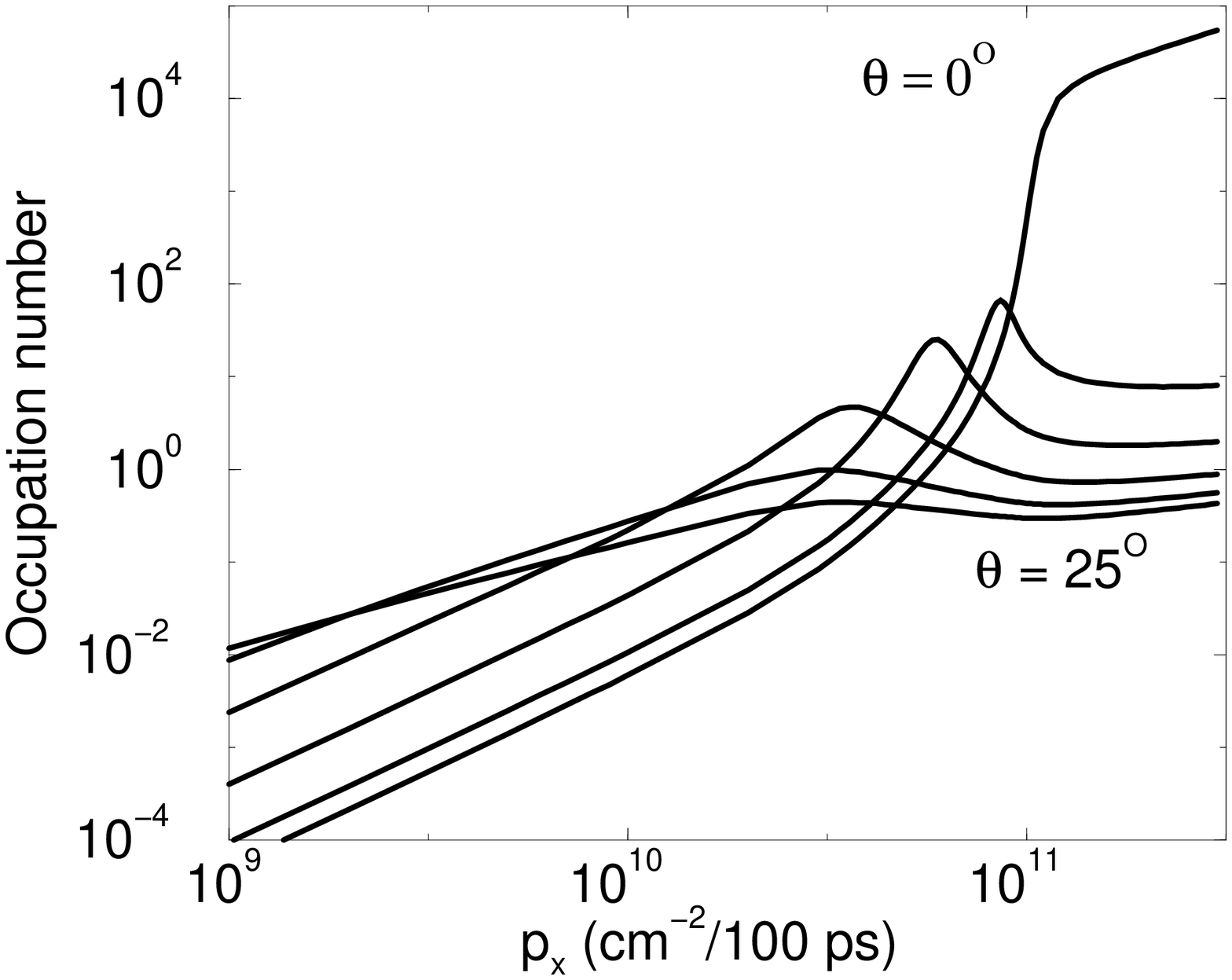,height=2.5in,width=2.5in}
}
\caption{Evolution of the occupancy of the modes corresponding to emission
angles $\theta = 0,5 \dots 25$ degrees (from right to left),
as pump-power is increased. $T_L=10K$,
$\Omega_P=10 meV$.}
\label{angles}
\end{figure}

As shown in Fig. \ref{peakPump10} for intermediate pump-powers the system does
not relax down to the ground state, but stimulation
(occupancy larger than one) 
is achieved in the scattering into intermediate
modes with $k \neq 0$. The stimulated modes will thus emit light at angles $\theta
\neq 0$. Fig. \ref{angles} presents the evolution of the occupation numbers for the
modes corresponding to different angles, as pump-power is increased. The
evolution of the occupancy at each angles matches the experimental results in
\cite{ring}.
A quadratic relation is predicted, followed by an
exponential growth that saturates at large powers. 
Stimulation can be observed, at polariton densities below the BEC,
at given angles,
which correspond to the modes where the polariton distribution peaks, as shown
in Fig. \ref{peakPump10}.
The saturation of the emission
at a given angle is related to the evolution of the polariton population that
"shifts downwards" in energy, due to the heating of the exciton reservoir. Thus,
the emission at a given angle saturates when the peak in the polariton
distribution turns to smaller angles, as pump-power is increased.

\section{Bose-Einstein Condensation of polaritons}

In the previous section it was shown that X-P scattering 
is efficient enough to
create a quasi-thermal population into the LPB. We will
quantify now in detail how the occupancy of the lowest energy polariton state
evolves when the polariton density is increased.

Above a critical pump-power, the continuous Bose-Einstein distribution
shown in Fig. \ref{peakPump10} saturates, and the additional
excitons added by the pump fall into the bottom of the LPB. Thus,
the lower polariton distribution becomes discontinuous, and 
an abrupt increase is achieved in the number of polaritons in the lowest
microscopic level, $N_0$. 
Above the threshold for BEC, 
the macroscopic occupancy of a single microscopic mode opens the possibility for
spontaneous symmetry breaking and the appearance of coherence in that mode. 
The interaction between the Bose-Einstein condensate and the low energy polariton
modes would lead then to the reconstruction of the polariton spectrum, as predicted
in the well known Bogoliubov approach
\cite{Moschkalenko}. 
These effects can modify the relaxation dynamics above threshold, so that our
results are not rigorous in this range. However we can predict the critical
density and temperature for BEC, and the most favorable conditions for its
observation.

\begin{figure}
\centerline{
\psfig{figure=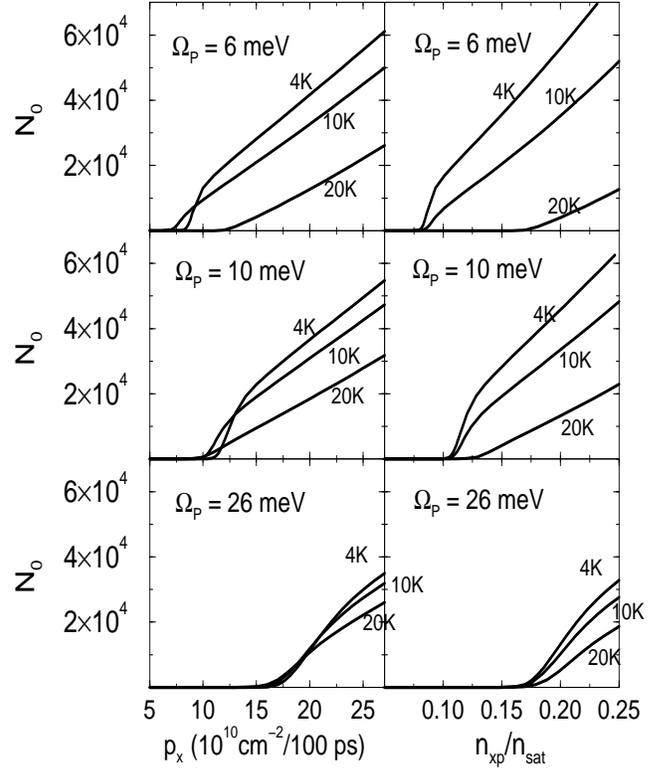,height=4.5in,width=3.3in}
}
\caption{Left: Evolution of the occupancy of the lowest microscopic polariton
level, as a function of the pump-power. For each polariton splitting we consider $T_L
= 4, 10, 20 K$. Right: Same as above, but now the occupancy of the lowest
microscopic level is plotted as a function of the steady-state exciton-polariton
density.}
\label{bec}
\end{figure}

Fig. \ref{bec} (left)
shows the evolution of $N_0$, as a function of $p_{x}$, for different lattice
temperatures and polariton splittings. 
For large $p_x$ above the critical pump, the relation between $N_0$ and $p_x$
becomes linear, and occupation numbers of the order of $5 \times 10^4$ 
polaritons can be reached
in the ground state. In order to interpret the evolution of the ground state
occupancy, it is interesting to plot $N_0$ (Fig. \ref{bec} (right)) as a function
of the steady-state exciton-polariton density, $n_{xp}$.
The abrupt increase in $N_0$ allows us to estimate a critical density
$n_{xp}^c$ for BEC,
which results to be around $0.1 n_{sat}$. 
However, the situation is now not so clear as in the case of
equilibrium BEC, where one expects that $N_0$ follows the relation (see
Appendix B):
\begin{equation}
\frac{N_0}{S} = n_{xp} - n^c_{xp} .
\label{eqBEC}
\end{equation}
That is, for densities above $n^c_{xp}$, all the excitons added to the system
should fall into the ground state. Fig. \ref{bec} (right) 
shows that, for large densities, the linear
relation is satisfied. However, close to $n_{xp}^c$, $N_0$ shows a superlinear
dependence on $n_{xp}$. This fact can be interpreted as follows. The equilibrium BEC
relation between $N_0$ and $n_{xp}$ would be given by the extrapolation of the
linear regime of the curves in Fig. \ref{bec} (right), towards $N_0 \approx 0$. 
By this method, we would
obtain the equilibrium $n_{xp}^c$,
which is smaller than the one we get from our complete nonequilibrium
calculation. This means that in a real microcavity, $n_{xp}^c$ is larger than the one
predicted by equilibrium BEC. Obviously, this is due to the radiative losses in
the LPB, and also to the fact that the system has to evolve towards the ground
state, following the steps described in the previous section.
Thus, the X-P scattering has to
overcome the radiative losses, and the exciton reservoir has to be heated, so that
relaxation into the ground state becomes efficient enough. 
The nonequilibrium critical density
has to be larger than the equilibrium one, for these conditions to be fulfilled.  

The transition from superlinear to linear dependence of $N_0$ as a function of
$p_x$, and $n_{xp}$, is reflected in the evolution of the exciton reservoir
temperature, $T_x$. Fig. \ref{tempEvol} shows that the heating of the exciton gas
slows down for the pump-power for which the linear growth of $N_0$ starts (see
Fig. \ref{bec}, left panel, $\Omega_P = 10 meV$). The same behavior in the
evolution of $T_x$ has been obtained for all the $\Omega_P$'s considered.  

We can also see in Fig. \ref{bec},
that for small polariton splittings,
smaller pump-powers and polariton densities are needed for BEC, 
because the
energy distance between the bottom of the LPB and the exciton reservoir is also
smaller. For large $\Omega_P$, the exciton gas reaches higher temperatures in
the steady-state regime, and the X-P scattering rates are slower
(thus, larger polariton densities are needed in order to get BEC). 
However, in
all cases considered in both figures, the critical density is well below
$n_{sat}$. 

\begin{figure}
\centerline{
\psfig{figure=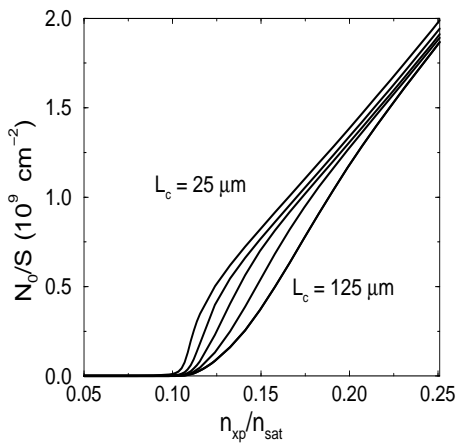,height=2.5in,width=2.5in}
}
\caption{Evolution of the density of polaritons in the lowest microscopic state,
as the exciton-polariton density is increased,
for different quantization lengths, $L_c
= 25,50 - 125 \mu m$, and $\Omega_P = 10 meV$, $T_L = 10 K$.}
\label{beclc}
\end{figure}

Up to now we have assumed $L_c = 50 \mu m$, of the order of typical spot
diameters. However, $L_c$ is not a well determined quantity in experiments.
Equilibrium BEC in 2-D systems predicts that both 
critical density and temperature depends only on the order of magnitude of $L_c$. In
order to check what happens in our nonequilibrium situation, we have calculated
the density of polaritons in the ground state, $N_0/S$, for different quantization
lengths (see Fig. \ref{beclc}). 
The occupancy of the lowest polariton mode has a more abrupt increase for
smaller $L_c$. However
$n_{xp}^c$ is of the order of $0.1 n_{sat}$ for all the quantization lengths
considered.

\section{Conclusions}

Our main conclusions are: 
{\it (i)} At large densities, relaxation after continuous
non-resonant excitation can be described by a model in which lower polaritons are
injected from a thermalized exciton reservoir by X-P scattering.
{\it (ii)} Our
model shows a linear relation between the energy of the maximum in the lower
polariton distribution and the exciton temperature or pump-power.
By increasing the exciton density, the heating of the exciton reservoir allows
the excitons to overcome the phase-space restrictions and fall into the LPB,
that becomes a polariton trap in reciprocal space. 
{\it (iii)}
Bose-Einstein Condensation is possible for densities well below $n_{sat}$
for CdTe microcavities. 
{\it (iv)} The most favorable conditions for BEC are 
small polariton splittings, due to the smaller 
difference between the bare exciton energy and the bottom of the LPB. 
{\it (v)} Our
results have a weak dependence on the quantization length, which is taken as the
excitation spot diameter.

Up to now, non-resonant 
experiments performed in GaAs microcavities confirm that the weak
coupling regime is reached before X-P scattering allows polaritons to relax towards
the ground state \cite{ring,Senellart}. Experimental evidence for the intermediate
stage in the evolution of the polariton distribution has been achieved in
a GaAs microcavity \cite{ring}, and the experimental results agree very well 
with the results presented in section III. In CdTe,
experimental evidence for stimulation towards the lower polariton states has been
achieved under pulsed non-resonant excitation.
The angle-resolved measurements after pulsed excitation in CdTe presented in
\cite{tesis} qualitatively agree with our results for the evolution of the
polariton distribution as pump-power is increased. 
However, the physical situation in
pulsed experiments is quite different than the one considered here, because the
system is not allowed to reach the steady-state in the relaxation process.
Continuous wave excitation seems to be a cleaner experimental situation for the
study of the evolution of the polaritons towards a quasi-equilibrium distribution.
Our model allows to predict not only
the possibility of BEC, but also quantitatively describes the evolution
of the system towards the ground state. 
Angle-resolved measurements under continuous non-resonant excitation would allow to
show the macroscopic occupancy of the $k=0$ mode, and also to monitor the evolution
towards BEC as pump-power is increased.

We are grateful to Michele Saba, Luis Vi\~na, Francesco Tassone, 
Ulrich R\"ossler, Antonio
Quattropani, and Paolo Schwendimann, for fruitful discussions. D. Porras
acknowledges the hospitality of the \'Ecole Polytechnique F\'ed\'eral de
Lausanne. 
Work supported in part by MEC (Spain) under contract Nº PB96-0085, and CAM
(Spain) under contract CAM 07N/0064/2001.

\appendix

\section{Calculation of the X-P scattering rates}

The calculation of the rates for the scattering 
of thermalized excitons into the LPB
was performed in \cite{Tassone}. Here we reproduce, for clarity, the main
steps, and calculate also the rate for {\it ionization} out of the LPB.

We consider first the process that injects polaritons. It is given by:
\begin{eqnarray}
W^{in}_k && n^2_x (1+N^{lp}_k) = \nonumber \\
&& \frac{4}{\hbar \pi^3} |M_k|^2 \rho_x^3 \int 
R(\epsilon_i^x,\epsilon_j^x,\epsilon_f^x,\epsilon_k^{lp}) \nonumber \\
&& \ \ \ \ \ N_i^x N_j^x (1+N_k^{lp}) d \epsilon_i^x d \epsilon_j^x  .
\label{cRatein}
\end{eqnarray}
$N_i^x$, $N_j^x$ are the occupation numbers for the initial excitons in the
reservoir, and $N_f^x$, $N_k^{lp}$ correspond to the final exciton and lower
polariton, respectively. 
The $N^x$ in the thermalized exciton reservoir follow the relation (\ref{MB}).
Due to their large density of sates we can neglect the
stimulation terms in the exciton reservoir ($N^x << 1$). 
Energy conservation is implicit in Eq. (\ref{cRatein}), so that $\epsilon_f =
\epsilon_i + \epsilon_j - \epsilon_k^{lp}$.
The function $R$ describes
the phase-space restrictions for the problem of exciton-exciton 
scattering in 2-D,
under the approximation of isotropic exciton distributions. It can be
expressed in terms of the corresponding in-plane momenta \cite{Tassone}:
\begin{eqnarray}
&&R(\epsilon_i^x,\epsilon_j^x,\epsilon_f^x,\epsilon_k^{lp}) = \nonumber \\ 
&& \int \frac{1}{2} dx  
\left(  \ \  
\left[ (k_i^x \! \! + \! \! k^{lp})^2 \! - \! x \right]
\left[ x \! \! - \! \! (k_i^x \! \! - \! \!  k^{lp})^2 \right] \ \ \times
\right. \nonumber \\
&&  \left. \left[ (k_j^x \! \! + \! \! k_f)^2 \! - \! \! x \right]
\left[ x \! \! - \! \! (k_j^x \! \! - \! \! k^f)^2 \right]
 \ \ \right) ^{-\frac{1}{2}} ,
\label{laerre}
\end{eqnarray}
where integration is limited to the interval where the argument of the square
root is positive. A great simplification is obtained when we take the limit
$k^{lp} << k^x$:
\begin{equation}
R(\epsilon_i^x,\epsilon_j^x,\epsilon_f^x,\epsilon_k^{lp}) = 
\frac{\pi}{8 \rho_x} 
\frac{1}{\sqrt{\epsilon_i^x \epsilon_j^x - 
\left(  \epsilon_i^{lp} /2  \right)^2
}} .
\label{laerresimple}
\end{equation}
If we insert this expression into (\ref{cRatein}), we obtain
\begin{eqnarray}
W^{in}_k =
&& \frac{2 \pi}{\hbar (k_B T_x)^2} |M_k|^2 \nonumber \\ 
&& \int 
\frac{1}{\sqrt{\epsilon_i^x \epsilon_j^x - 
\left(  \epsilon_i^{lp} /2  \right)^2}}
e^{-(\epsilon_i^x +\epsilon_j^x) / k_B T_x } 
d \epsilon_i^x d \epsilon_j^x ,
\label{cRateinsimple}
\end{eqnarray}
which after integration in energy gives the expression reported in Eq. 
(\ref{rates}). 

The calculation of the rates for scattering out of the LPB 
follows the same way:

\begin{eqnarray}
&& W^{out}_k n_x N^{lp}_k = \nonumber \\ 
&& \frac{4}{\hbar \pi^3} |M_k|^2 \rho_x^3 \int 
R(\epsilon_f^x,\epsilon_k^{lp},\epsilon_i^x,\epsilon_j^{lp})
N_f^x N_k^{lp} d \epsilon_i^x d \epsilon_j^x .
\label{cRateout}
\end{eqnarray}
The factor $R$ is again given by (\ref{laerresimple}), because phase-space
restrictions are the same as in the previous case,
\begin{eqnarray}
W^{out}_k =
&& \frac{\rho_x}{\hbar \pi (k_B T_x)^2} |M_k|^2 \nonumber \\ 
&& \int 
\frac{1}{\sqrt{\epsilon_i^x \epsilon_j^x - 
\left(  \epsilon_i^{lp} /2  \right)^2}}
e^{-\epsilon_f^x / k_B T_x } 
d \epsilon_i^x d \epsilon_j^x .
\label{cRateoutsimple}
\end{eqnarray}
$\epsilon_f^x$ can be expressed in terms of $\epsilon_i^x$, $\epsilon_j^x$,
and the expression in Eq. (\ref{rates}) is obtained.

\section{Bose-Einstein Condensation in two dimensions}

Let us consider an ensemble of free bosons that follow a Bose-Einstein
distribution with chemical potential $\mu<0$, and finite temperature $T$.
The density of particles can be evaluated as:

\begin{equation}
n(\mu,T)=\frac{1}{2 \pi} \int_{0}^{\infty} \rho (\epsilon)
\frac{d \epsilon}{e^{(\epsilon-\mu)/k_B T} -1} .
\label{BE}
\end{equation}

In 3-D, $\rho(\epsilon) \propto \sqrt{\epsilon}$, and the continuous Bose-Einstein
distribution with $\mu=0$ gives a finite critical density $n_c(T)$. For densities
$n>n_c(T)$, all the extra particles fall into the ground state. 

In 2-D,
$\rho = m/\hbar^2$ (constant), and the integration in Eq. (\ref{BE}) diverges as
$\mu \rightarrow 0$. For an infinite system, BEC is thus not possible at
finite temperature. For a finite system, we can introduce a quantization length,
$L_c$, such that the integration in Eq. (\ref{BE}) has a low energy cut off,
$\epsilon_c$:
\begin{equation}
n(\mu,T)=\frac{m}{2 \pi \hbar^2} \int_{\epsilon_c}^{\infty}
\frac{d \epsilon}{e^{(\epsilon-\mu)/k_B T} -1} .
\label{BE2D}
\end{equation}
For the case $\epsilon_c/k_B T << 1$, Eq. (\ref{BE2D}) gives a very simple result
for the maximum density that the 2-D B-E distribution can accommodate:
\begin{equation}
n_c(T) = - \frac{m k_B T}{2 \pi \hbar^2} 
log \left( \frac{\epsilon_c}{k_B T} \right) .
\label{critical}
\end{equation}
Whenever condition $n>n_c(T)$ is satisfied, the extra particles added to the
system fall into the ground state:
\begin{equation}
\frac{N_0(T)}{S} = n - n_c(T) .
\label{N0}
\end{equation}
For free particles, $\epsilon_c \propto \L_c^{-2}$, and Eq. (\ref{BE2D}) allows to
predict a weak logarithmic dependence of the critical density on the quantization
length. This result qualitatively explains the weak dependence of the critical
density on $L_c$,
obtained in the case of nonequilibrium BEC discussed above.

\widetext


\begin{references}
\bibitem{note} The term Rabi splitting ($\Omega_R$), is also used for the
polariton splitting in this context.
We have changed here the traditional terminology in order to avoid confusion
with the case of splitting in a two-level fermionic system, where the term Rabi
splitting was initially introduced.
\end{references}
\end{document}